\documentclass[aps,prb,groupedaddress,floats,reprint]{revtex4-2}
\usepackage[german,english]{babel}

\usepackage{amsfonts,graphicx,soul,natbib,mathrsfs}
\usepackage[dvipsnames]{xcolor}

\usepackage[normalem]{ulem}  

\usepackage[colorlinks=true]{hyperref}

\usepackage{epstopdf}
\usepackage{textcomp}
\usepackage{xcolor}
\usepackage{soul}
\usepackage{physics}
\usepackage{graphicx}
\usepackage{lipsum}
\usepackage{comment}
\usepackage{amssymb}
\usepackage{array}

\newcommand{\expv}[1]{\left\langle #1 \right\rangle}

\usepackage{bm}

\usepackage{amsmath,amssymb,amsfonts}
\usepackage{fancyhdr}
\usepackage{lipsum}
\usepackage{subcaption}
\usepackage{multirow}
\usepackage{dcolumn}
\usepackage{bm}

\usepackage{titlesec}
\usepackage{hyperref}

\titleclass{\subsubsubsection}{straight}[\subsection]

\newcounter{subsubsubsection}[subsubsection]
\renewcommand\thesubsubsubsection{\thesubsubsection.\arabic{subsubsubsection}}

\titleformat{\subsubsubsection}
  {\normalfont\normalsize\bfseries}{\thesubsubsubsection}{1em}{}
\titlespacing*{\subsubsubsection}
{0pt}{3.25ex plus 1ex minus .2ex}{1.5ex plus .2ex}
\makeatletter
\renewcommand\paragraph{\@startsection{paragraph}{5}{\z@}%
  {3.25ex \@plus1ex \@minus.2ex}%
  {-1em}%
  {\normalfont\normalsize\bfseries}}
\renewcommand\subparagraph{\@startsection{subparagraph}{6}{\parindent}%
  {3.25ex \@plus1ex \@minus .2ex}%
  {-1em}%
  {\normalfont\normalsize\bfseries}}
\def\toclevel@subsubsubsection{4}
\def\toclevel@paragraph{5}
\def\toclevel@paragraph{6}
\def\l@subsubsubsection{\@dottedtocline{4}{7em}{4em}}
\def\l@paragraph{\@dottedtocline{5}{10em}{5em}}
\def\l@subparagraph{\@dottedtocline{6}{14em}{6em}}
\makeatother

\begin{document}

\title{Elastic scattering of Laguerre - Gaussian electron packets on atoms}

\author{N. Sheremet}%
\affiliation{School of Physics and Engineering,
ITMO University, 197101 St. Petersburg, Russia}%

\author{A. Chaikovskaia}%
\affiliation{School of Physics and Engineering,
ITMO University, 197101 St. Petersburg, Russia}%

\author{D. Grosman}%
\affiliation{School of Physics and Engineering,
ITMO University, 197101 St. Petersburg, Russia}%

\author{D. Karlovets}%
\affiliation{School of Physics and Engineering,
ITMO University, 197101 St. Petersburg, Russia}

\begin{abstract}
  We explore 
 elastic scattering of non-relativistic electrons in the form of 
 standard Laguerre-Gaussian (sLG) and elegant Laguerre - Gaussian (eLG) packets on atomic targets in the generalized Born approximation 
 and compare 
 these results 
 to the reference with 
 Bessel - Gaussian (BG) packets. 
Scattering by hydrogen-like, iron, silver, and golden targets is considered. 
The incident electron carries a nonzero 
orbital angular momentum, while sLG and eLG packets 
have a definite radial quantum number $n$ as well.
In scattering of 
 sLG and eLG wave packets by a macroscopic target sensitivity of the average cross section to the orbital angular momentum is observed, which is absent for 
 BG packets. We highlight the opportunity to employ the differences in the experimental scattering results for the revelation of the properties of incident twisted electron wave packets. 
\end{abstract}

\maketitle

\section{INTRODUCTION}
There is a 
class of the Schr\"{o}dinger equation solutions that describe vortex, or twisted, beams, i.e. those carrying an orbital angular momentum (OAM). Laguerre-Gaussian states, widely recognized both in laser optics~\cite{ IEEE_LG1966, siegman1986lasers, Allen1992} and in physics of electron beams \cite{bliokh2012, BLIOKH20171}, are 
square-integrable solutions of this kind with 
a 
well-defined projection of the OAM $l=0, \pm 1, \pm 2, \dots$, and a 
radial quantum number $n=0,1,2, \dots$. There are two varieties of such beams - 
standard Laguerre-Gaussian (sLG) and elegant Laguerre-Gaussian (eLG) wave packets~\cite{siegman1986lasers, Kostenbauder97,LU2000, NJP2021} - both are exact solutions of the Schr\"{o}dinger equation and are equally utilized in this paper. A specific feature of 
eLG modes is that, though these states form a complete set of solutions to the free-space Schr\"{o}dinger equation, they are not orthogonal~\cite{Siegman73, siegman1986lasers}. Moreover, sLG packets spatial distribution have $n+1$ peaks, whereas for eLG packets, the second peak becomes pronounced at large values of $n$. 
Alternate vortex solutions are monochromatic Bessel beams
~\cite{BagrovGitman,Berry1979}, which are characterized by longitudinal wave number $k_z=p_z/\hbar$, transverse (radial) wave number $\varkappa=p_{\perp}/\hbar$, and projection of OAM $l$. However, such states spread infinitely and are non–square–integrable solutions. In this work we employ their convolution with a function of the Gaussian form, and thus Bessel-Gaussian beams are considered.

In this paper, we analyze 
elastic scattering of 
non-relativistic electron wave packets by atomic targets in the 
generalized Born approximation. The incident electron is 
in the form of sLG and eLG wave packets, and the targets are 
hydrogen, iron, silver, and gold atoms. 
We explore how the shape of the spatial profile affects 
the magnitude of the average 
cross section and the number of scattering events. We benchmark 
our results with the scattering model for BG electron wave packets, implemented in Ref.~\cite{Karlovets2017}. Thus, by utilizing the LG packets our results give a more realistic description of the scattering, in particular, demonstrating that the cross section of the macroscopic target may  depend substantially on the OAM, in contrast to the simplified BG case. 
Also, the best experimental motivation for studies with various forms of wave packets comes from the optics. There, flexible beam shaping and, in particular, generation of LG optical beams of different types began with the advent of spatial light modulators. It is frequently discovered that the distinguishing properties of different light modes enable better applications in some fields. For instance, it was demonstrated that elegant Gaussian beams enhance trapping forces in optical micro manipulation~\cite{alpmann2015elegant}.
 
In practice, twisted electron beams are produced by phase plates~\cite{Uchida10}, 
diffracting holograms~\cite{Benjamin11, Grillo14,Verbeeck10}, and in 
scanning electron microscopes~\cite{Verbeeck10, Grillo15}; 
the achieved values of the OAM projection are as high as $\hbar l = 1000\hbar$~\cite{l1000}. 
Electron beams find their applications in various fields~\cite{Verbeeck10, BLIOKH20171}, particularly, in 
manipulation of particles~\cite{Verbeeck12}, electron trapping~\cite{Pena22}, and strong-field chiral imaging~\cite{Planas2022}.
Active theoretical study of massive vortex beams goes in parallel, let us mention several relevant results.
Elastic and inelastic scatterings of vortex electrons parameterized as Bessel beams on atoms were 
investigated in Refs.~\cite{VanBoxem2014,VanBoxem2015}. 
There is a number of reports on scattering of 
Gaussian wave packets~\cite{Karlovets2015}, 
twisted BG electron wave packets~\cite{Karlovets2017}, 
Schr\"{o}dinger cat states~\cite{KarlovetsShred2017}, and 
Airy electron wave packets~\cite{Grosman2023} on atoms. 
Twisted electrons' scattering beyond the Born approximation was 
considered in Ref.~\cite{Kosheleva2018}. 
Additionally, the processes of atom 
ionization by twisted electron beams ~\cite{Harris_2019, Plumadore_2020, Dhankhar_2020} and angular momentum exchange ~\cite{LLoyd2012_Dichroism, Lloyd2012} were 
explored. 
Scattering of twisted beams by molecules was 
studied in Refs.~\cite{Maiorova18, harris2023, Dhankhar22}.

\begin{figure*}
 \begin{subfigure}{0.32\textwidth}
     \includegraphics[width=\textwidth]{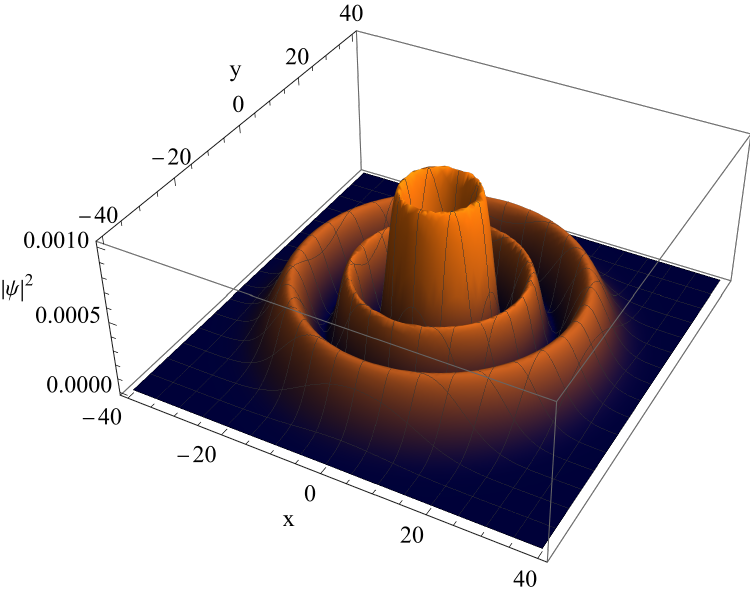}
     \caption{}
     \label{fig:ProfLGS}
 \end{subfigure}
 \hfill
 \begin{subfigure}{0.32\textwidth}
     \includegraphics[width=\textwidth]{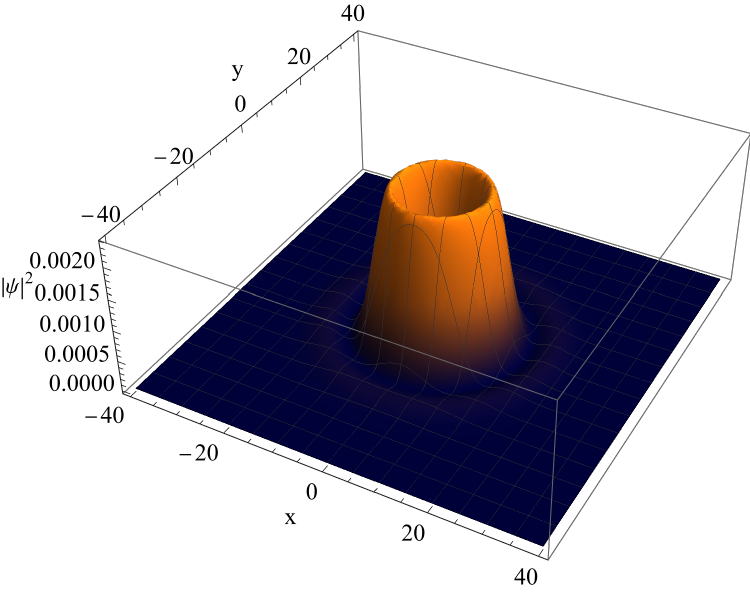}
     \caption{}
     \label{fig:ProfLGE}
 \end{subfigure}
 \hfill
 \begin{subfigure}{0.32\textwidth}
     \includegraphics[width=\textwidth]{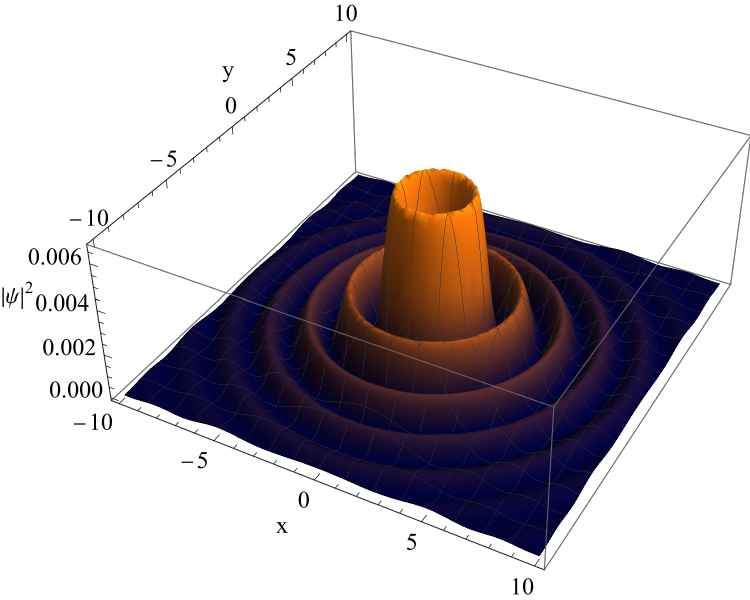}
     \caption{}
     \label{fig:ProfB}
 \end{subfigure}
 \caption{Spatial distribution of the wave
packets with $l=2$: a) Standard Laguerre - Gaussian packet: $n=2$ ; b) Elegant Laguerre - Gaussian packet: $n=2$; and c) Bessel - Gaussian packet.}
 \label{fig:Prof}
\end{figure*}

The structure of the paper is as follows. First, we provide the 
necessary theoretical background. In Sec.~\ref{Sec:2}, we introduce various 
twisted wave packets. The theoretical formalism for 
scattering of twisted electron  packets by a single atom is presented in Sec.~\ref{Sec:3A}. Sec.~\ref{Sec:3D} is devoted to 
scattering of 
electron packets on a macroscopic target. In Secs.~\ref{Sec:3B} and~\ref{Sec:3C},  we consider two types of the potential field off 
which the incident electron wave packets scatter. 
The options with hydrogen potential and superposition of three Yukawa terms are investigated. 
Then, in Sec.~\ref{sec:4}, we discuss different scenarios of scattering of twisted electron 
wave packets 
and illustrate the number of events and average 
cross sections for different values of the OAM and radial quantum number. The summary is provided 
in Sec.~\ref{Sec:5}.
A system of units with $\hbar=c = 1$ is further used.

\section{TWISTED WAVE PACKETS}
\label{Sec:2}

In this section, we consider different types 
of well-localized twisted packets: standard Laguerre - Gaussian, elegant Laguerre - Gaussian, and Bessel - Gaussian wave packets. All these wave packets 
have a defined projection of orbital angular momentum (OAM) and a symmetric spatial profile. 
\subsection{Standard Laguerre - Gaussian wave packet}
\label{Sec:2A}
An 
electron sLG state is obtained as an exact solution of the Schr\"{o}dinger equation in the cylindrical coordinates $\bm r=(\bm{\rho},z)=(\rho,\varphi_r,z)$~\cite{siegman1986lasers, NJP2021}:
\begin{equation}
\label{eq:LGC}
       \Psi_{sLG}(\bm{\rho},z,t) = \Psi_{\perp}^{sLG}(\bm{\rho},t) \exp\left(ik_z z-i\frac{k_z^2}{2m}t\right)
\end{equation}
with the transverse component in the coordinate representation being
\begin{equation}
\label{eq:sLGCt}
\begin{aligned}
\Psi&_{\perp}^{sLG}(\bm{\rho},t)= N_{sLG} \frac{\rho^{|l|}}{\sigma_{\perp}^{|l|+1}(t)}
 L_{n}^{|l|}\left(\frac{\rho^2}{\sigma^2_{\perp}(t)}\right) \\
&\times \exp\left[il\varphi_r-i(2n+|l|+1)\arctan(t/t_d)\right] \\
&\times \exp\left(-\frac{\rho^2}{2\sigma_{\perp}^2(t)}(1-it/t_d)\right),
\end{aligned}
\end{equation}
where $L_{n}^{|l|}$ are 
generalized Laguerre polynomials, $l=0,\pm1,\pm2, \dots$ is the OAM, $n=0,1,2,\dots$ is the radial quantum number that indicates 
the number of rings of the sLG wave packet, $\sigma_{\perp}$ is the packet 
width, 
and $N_{sLG} = \sqrt{\frac{n!}{\pi(|l|+n)!}}$ is determined from the normalization condition
$\int d^2{\bm \rho}|\Psi_{\perp}^{sLG}(\bm{\rho},t)|^2=1$.
The longitudinal component of the wave function \eqref{eq:LGC} 
is a standard plane wave
and, from here on, we focus only on the transverse component of the packets' wave functions.
We also neglect the packets' dispersion ($t=0$), and therefore, 
the wave function \eqref{eq:sLGCt} becomes 
\begin{equation}
\label{eq:sLGC}
\Psi_{\perp}^{sLG}(\bm{\rho})= N_{sLG} \frac{\rho^{|l|}}{\sigma_{\perp}^{|l|+1}} L_{n}^{|l|} \left(\frac{\rho^2}{\sigma_{\perp}^2}\right)\exp(il\varphi-\frac{\rho^2}{2\sigma_{\perp}^2}).
\end{equation}

It is convenient to use 
the Fourier transform and 
write the wave function \eqref{eq:sLGC} in the momentum representation
\begin{equation}
\label{eq:LGI}
     \Psi_{\perp}^{sLG}(\bm{\rho}) =\int \frac{d^2\bm{k}_{\perp}}{(2\pi)^2}\Phi_{\perp}^{LG}(\bm{k}_{\perp}) e^{i \bm{k}_{\perp} \bm{\rho}},
\end{equation}
where
\begin{equation}
\label{eq:sLGI}
	\Phi_{\perp}^{sLG}(\bm{k}_{\perp}) = N_{sLG} \frac{k_\perp^{|l|}}{{\sigma_{\varkappa}^{|l|+1}}} L_{n}^{|l|} \left(\frac{k_\perp^2}{\sigma_{\varkappa}^2}\right)\exp(il\varphi_{k}-\frac{k_\perp^2}{2\sigma_{\varkappa}^2}).
\end{equation}
Here $\sigma_{\varkappa}=1/\sigma_{\perp}$ is the packet width in the momentum representation.

The probability density  $|\Psi_{\perp}^{sLG}(\bm{\rho})|^2$ in the coordinate representation is shown in Fig.~(\ref{fig:Prof}, a) for $l=2$ and $n=2$.

\subsection{Elegant Laguerre - Gaussian wave packet}
\label{Sec:2B}
\begin{figure*}
 \begin{subfigure}{0.32\textwidth}
     \includegraphics[width=\textwidth]{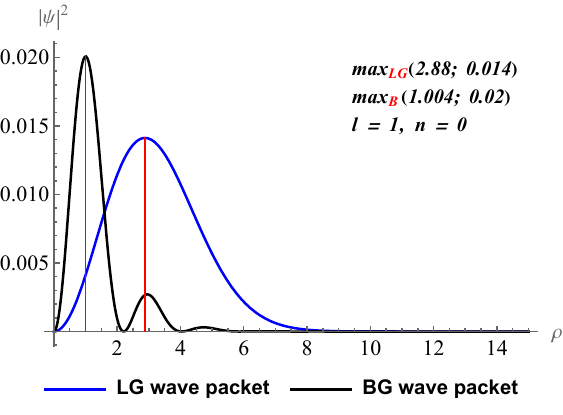}
     \caption{}
 \end{subfigure}
 \hfill
 \begin{subfigure}{0.32\textwidth}
     \includegraphics[width=\textwidth]{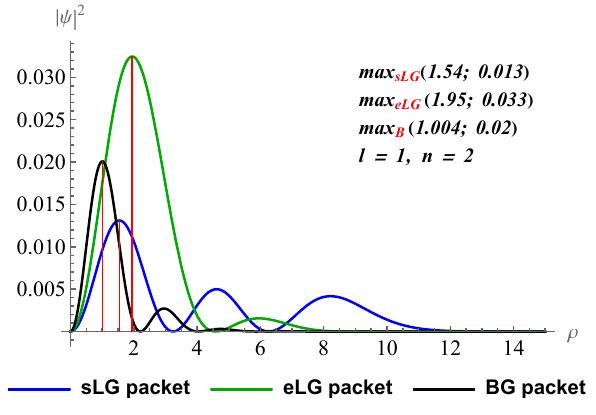}
     \caption{}
 \end{subfigure}
 \hfill
 \begin{subfigure}{0.32\textwidth}
     \includegraphics[width=\textwidth]{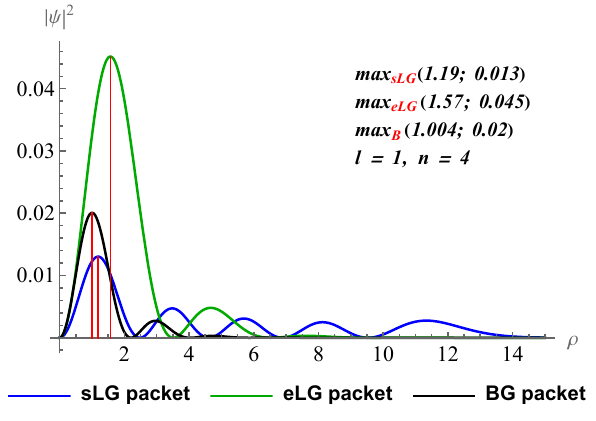}
     \caption{}
 \end{subfigure}
 
  \caption{Spatial
  distribution of 
  sLG (blue) and 
  eLG (green) wave packets for 
  different values of $n$: $n = 0$ (a); $n = 1$ (b); $n = 2$ (c), and BG (black) wave packet for $l = 1$. The packets width $\sigma_{\perp}=5/\varkappa$} is taken.
 \label{fig:Psen}
\end{figure*}

\begin{figure*}
 \begin{subfigure}{0.32\textwidth}
     \includegraphics[width=\textwidth]{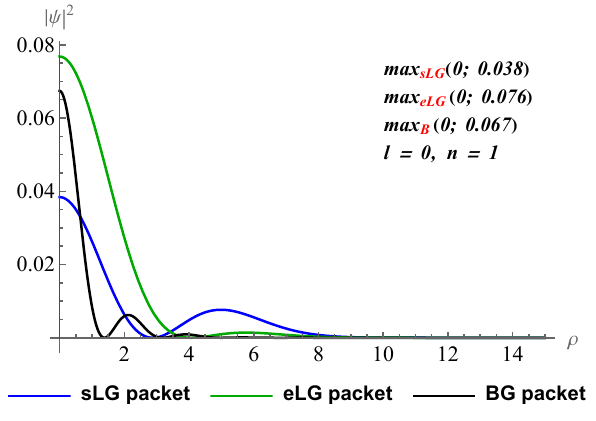}
     \caption{}
 \end{subfigure}
 \hfill
 \begin{subfigure}{0.32\textwidth}
     \includegraphics[width=\textwidth]{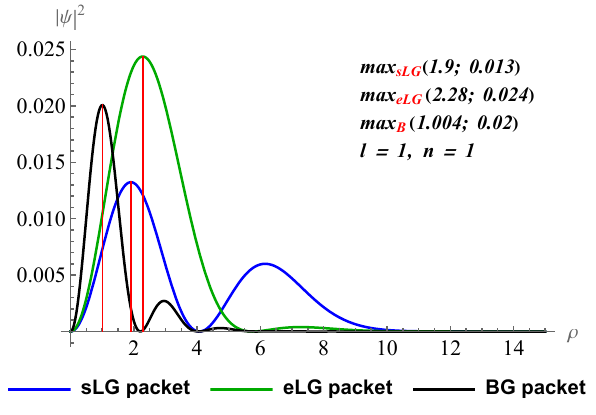}
     \caption{}
 \end{subfigure}
 \hfill
 \begin{subfigure}{0.32\textwidth}
     \includegraphics[width=\textwidth]{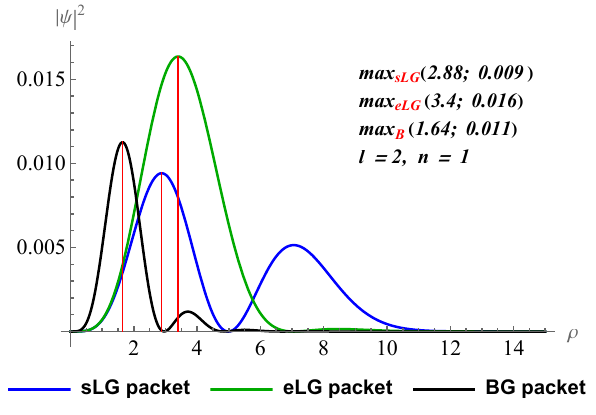}
     \caption{}
 \end{subfigure}
 
  \caption{Spatial distribution of 
  BG (black), sLG (blue), and 
  eLG (green) wave packets for 
  different values of $l$: $l = 0$ (a); $l = 1$ (b); $l = 2$ (c); $n = 1$.The packets width $\sigma_{\perp}=5/\varkappa$} is taken.
 \label{fig:PseL}
\end{figure*}

There is another exact solution of 
the Schr\"{o}dinger equation: 
an eLG state~\cite{siegman1986lasers, Siegman73,Zauderer86,NJP2021}, whose 
transverse component 
in the coordinate representation has the following form at $t=0$ 
\begin{equation}
\label{eq:eLGC}
    \Psi_{\perp}^{eLG}(\bm{\rho}) = N_{eLG}\frac{\rho^{|l|}}{{\sigma_{\perp}^{|l|+1}}} L_{n}^{|l|}  \left(\frac{\rho^2}{2\sigma_{\perp}^{2}}\right)\exp(il\varphi-\frac{\rho^2}{2\sigma_{\perp}^2}),
\end{equation}
where $N_{eLG}=\frac{2^{n} n!}{\sqrt{\pi (2n+|l|)!}}$. 
In the momentum representation, this function is given by
\begin{equation}
\label{eq:eLGI}
\begin{aligned}
    \Phi_{\perp}^{eLG}(\mathbf{k_\perp}) &=N_{eLG}(-1)^n\frac{k_\perp^{|l|}}{{\sigma_{\varkappa}^{|l|+1}}} \\
    &\times L_{n}^{|l|+n}  \left(\frac{k_\perp^2}{2\sigma_{\varkappa}^2}\right)\exp(il\varphi_{k}-\frac{k_\perp^2}{2\sigma_{\varkappa}^2}).
\end{aligned}
\end{equation}
Unlike the sLG states \eqref{eq:LGI}, the set of these functions is not orthogonal. Moreover, the spatial distribution of an 
eLG packet, for $n > 1$, has a different form~\cite{LU2000, Saghafi09}, as depicted in Fig.~(\ref{fig:Prof},b). 
Although for the sLG wave packet in Fig.~(\ref{fig:Prof}, a), there are three rings for $n=2$, in Fig.~(\ref{fig:Prof},  b) for the eLG packet one apparent ring and another barely noticeable one are observed. 
This is caused by the exponential term  $\exp(-\frac{\rho^2}{2\sigma_{\perp}^2})$ in Eq.~\eqref{eq:eLGC}. If we equate 
the arguments of Laguerre polynomials for the
sLG and eLG packets by replacing $2\sigma_{\perp}$ with 
$\sigma_{\perp}$ in Eq.~\eqref{eq:eLGC}, then the exponent there will decay 
twice as fast as the one in the sLG wave function \eqref{eq:sLGC}. Therefore, for eLG wave packets, higher-order rings are strongly suppressed, as can be seen in Fig.~\ref{fig:Psen}.
Moreover, if we consider the root-mean-square (rms) radii of LG packets, which are proportional to $n$ and $l$ \cite{NJP2021}, then for the eLG packet it will also be suppressed relatively to the rms radius of the sLG packet.

In Fig.~\ref{fig:Psen} 
we present the spatial profiles of 
sLG (blue line) and eLG (green line) wave packets with $l = 1$ for different values of the radial quantum number $n$. When $n=0$, the number of sLG and eLG rings coincides, and their spatial distribution is essentially the same, then we will treat and dub these packets as general Laguerre-Gaussian (LG) packets. 
With an increase of $n$, 
the magnitude of the first peak of the eLG packet becomes larger.
The second peak becomes more noticeable due to the narrowing of the first peak. On the other hand, the magnitude 
of the first peak of the sLG wave packet changes not as much with the addition of new rings, but the first ring becomes narrower. 
These simple observations already indicate that the number of events and the average
cross sections for sLG and eLG wave packets on the atomic targets for $n > 0$ will be different.

There is a substantial dependence of the spatial distribution of the
electron sLG and eLG wave packets 
on the 
OAM value $l$ when $n \neq 0$, as 
presented in Fig.~\ref{fig:PseL} for $n=1$. 
The larger $l$ is, the smaller the magnitude of the first peak becomes.
Besides, 
the first ring of the eLG wave packet with $l \neq 0$ is thicker, and its width grows faster with increasing $l$ than that of  
the sLG wave packet ring.
We will explore how 
these factors affect the 
scattering processes in detail in Section~\ref{sec:4}.

In experiments, when an electron beam profile is taken, both several noticeable rings~\cite{Verbeeck10,Kotlyar06,Wang16,Rumala15} and a more pronounced single rings~\cite{Benjamin11, Larocque18} can be observed.
Also, many recent theoretical works are focused on 
eLG wave packets and their applications~\cite{MABENA2024, Mylnikov:20,Jolly:22,XIA2016}. Optical eLG beams have found experimental application, for instance, in optical micro manipulation. 
Hence, we find it instructive to include the eLG wave packets in our study: both for the sake of completeness and to prompt the experimental interest.

\subsection{Bessel - Gaussian wave packet}
\label{Sec:2C}
A pure 
Bessel state is a non-square-integrable solution. Here, we consider a beam with a realistic finite spatial profile, a 
 BG wave packet, see the details in Ref.~\cite{Karlovets2017}. In the momentum representation the Bessel wave function is convoluted with the weight function over the absolute value of the transverse momentum, $|\bm{k}_{\perp}|=\varkappa$, so the Fourier transform of the BG wave packet has the form
\begin{equation} 
\label{eq:PsiB}
\begin{aligned}
    \Phi_{\perp}^{BG}(\bm{k}_{\perp})&=\int\limits_0^{\infty}   \Phi_{\perp}^{B}(\bm{k}_{\perp})g_{\varkappa_0 \sigma_{\varkappa}}(\varkappa)d\varkappa \\
    &= (-i)^l \frac{e^{i l \varphi_k}}{\sqrt{2\pi k_{\perp}}} g_{\varkappa_0 \sigma_{\varkappa}}(k_{\perp}).
    \end{aligned}
\end{equation}
Here $\Phi_{\perp}^{B}(\bm{k}_{\perp})$ is a Bessel function in the momentum representation
\begin{equation}
    \Phi_{\perp}^{B}(\bm{k}_{\perp})=(-i)^l \frac{e^{i l \varphi_k}}{\sqrt{2\pi}}\frac{\delta (k_{\perp}-\varkappa)}{\sqrt{k_{\perp}}}
\end{equation}
with $\varkappa=\sqrt{2m_e \varepsilon-k_z^2}$, and an opening angle is defined as follows \begin{equation}
\tan{\theta_k} = \varkappa/k_z.
\end{equation}
The function $ g_{\varkappa_0 \sigma_{\varkappa}}(\varkappa)$ in Eq.~\eqref{eq:PsiB} is the Gaussian weight function:
\begin{equation}
     g_{\varkappa_0 \sigma_{\varkappa}}(\varkappa)= C e^{-(\varkappa-\varkappa_0)^2/(2 \sigma_{\varkappa}^2)},
\end{equation}
which is peaked around $\varkappa_0$, defining the average value of the transverse momentum $\langle k_{\perp}\rangle=\varkappa_0$, and has the width $\sigma_{\varkappa}$. The constant $C$ is determined from the normalization condition $\int|  g_{\varkappa_0 \sigma_{\varkappa}} (\varkappa)|^2 d\varkappa=1$. 

The probability density of a BG wave packet is shown in Fig.~(\ref{fig:Prof}, c). In Fig.~\ref{fig:Psen}, we can see that for fixed $l$ as $n$ increases, the sLG and eLG packets become more and more different from each other. In the same time, with growth of $n$, the value of the first peak of the sLG wave packet diminishes gradually, the number of the rings increases, and more similarity between the distributions of the sLG and BG packets gets through. However, the magnitude of the first 
maximum of the eLG packet grows considerably, and the 
spatial distributions of 
eLG and BG packets start to diverge more noticeably. 

As shown in Fig.~\ref{fig:PseL}, while $n$ is fixed,
the magnitude of the first maxima drops with growth of $l$ for the packets of all three types. When $l=0$, the spatial distributions have 
a Gaussian shape. Besides, 
the magnitude of the first maximum of the BG wave packet is slightly higher than that of the sLG packet, being closer to the value of the eLG packet.
However, for $l > 0$, the first ring of the BG wave packet has a radius smaller than that of the sLG and eLG wave packets.
Therefore, we can predict that during a 
central collision of an 
electron BG wave packet with an atom, most of it will hit the target, and 
its number of events will be higher than those of 
sLG and eLG wave packets.
At the same time, during a 
non-central collision, 
scattering processes strongly depend on the impact parameter $b$, and we postpone the particular discussion of this case to 
Sec.~\ref{sec:4}.
We also notice that the magnitude of an eLG packet is greater than that of BG and sLG packets.
The influence of all these factors on scattering processes is discussed in detail in Sec.~\ref{sec:4}.
 
\section{Scattering in Born approximation}
\label{Sec:3}
In real situation an electron is spatially localized in all three dimensions.
However, in experimentally relevant circumstances, the size of the packet in the direction of motion, $\sigma_z$, is usually larger than the field's characteristic radius of action $a$. Hence, in the following 
we consider 
wave packets whose 
dispersions 
are much smaller than the mean values of the longitudinal momentum, $p_i=\langle k_z \rangle$, of the incoming electron 
\begin{equation}
\label{disp}
         \Delta k_z \sim \frac{1}{\sigma_z} \ll p_i, \; \Delta k_x = \Delta k_y \sim \frac{1}{\sigma_{\perp}} \ll p_i,
\end{equation}
where $\sigma_{\perp}$ and $\sigma_z$ are the average 
transverse and longitudinal sizes of the electron wave packets. 
Here we also assumed that during the collision, the dispersion of the packet in the transverse plane is negligible:

\begin{equation}
\label{eq}
  a\ll\sigma_z\ll p_i/(\varkappa_0\sigma_{\varkappa}).  
\end{equation}
\subsection{Scattering on a single atom}
\label{Sec:3A}
We start the discussion with a single potential $U(r)$, $r=|\bm{r}|$ located
at a distance 
from the center of the electron wave
packet equal to the impact parameter $b$. With assumptions \eqref{disp},\eqref{eq}, we can follow the derivations 
of Ref.~\cite{Karlovets2015} and write down the 
expression for the 
number of scattering events 
\begin{equation}
 \label{eq:Nu}
  \frac{d\nu}{d\Omega} =\frac{N_e}{\cos{\theta_k}}|F(\bm{Q},\bm{b})|^2,
  \end{equation}
where $N_e$ is the number of electrons in the incident beam, and $F(\bm{Q},\bm{b})$ is the scattering amplitude of the wave packet given by
  \begin{equation}
  \label{eq:F}
  \begin{aligned}
  &F(\bm{Q},\bm{b}) = \int f(\bm{Q}-\bm{k}_{\perp})\Phi_{\perp}(\bm{k}_\perp)e^{i\bm{k}_\perp \bm{b}}\frac{d^2\bm{k_{\perp}}}{2\pi}, \\
 & \bm{Q}=\bm{p}_f-\bm{p}_i \\
 &=(p_f \sin{\theta}\cos{\varphi},p_f \sin{\theta}\sin{\varphi, p_f \cos{\theta}-p_i}).
  \end{aligned}
\end{equation}
Here, $\bm{Q}$ quantifies the momentum transfer from the initial particle with $\bm{p}_i=\expv{\bm{k}}=(0,0,p_i)$ to the final particle with $\bm{p}_f$, $p_f=\sqrt{p_i^2+\varkappa_0^2}$, $\bm{b} = (b_x,b_y,0)$ is the impact parameter that 
indicates the location of the target with respect 
to the incident beam and vanishes when we consider central collisions. $f(\bm{Q}-\bm{k}_{\perp})$ is the 
scattering amplitude of 
the incident plane wave 
determined by the potential field $U(r)$ 
\begin{equation}
\label{eq:f}
\begin{aligned}
     f&(\bm{q})=-\frac{m_e}{2\pi} \int U(r) e^{-i\bm{q}\bm{r}}d^3r, \\ \bm{q}&=\bm{p}_f-\bm{p}_i.
\end{aligned}
\end{equation}
Then, the scattering amplitude \eqref{eq:F} can be written as
\begin{equation}
\label{eq:FU}
   F(\bm{Q},\bm{b}) = -\frac{m_e}{2\pi}\int U(r) \Psi_{\perp}(\bm{r}+\bm{b})e^{-i\bm{Qr}}d\bm{r},
\end{equation}
where
\begin{equation}
    \Psi_{\perp}(\bm{r}+\bm{b})=\int e^{i(\bm{r}+\bm{b})\bm{k}_{\perp}}\Phi_{\perp}(k_{\perp})\frac{d^2\bm{k}_{\perp}}{2\pi}.
\end{equation}
If we consider only the azimuthal part of Eq.~\eqref{eq:FU}, for $\bm{b}=0$, and for any wave packet 
it will be equal to
\begin{equation}
   \int \limits_0^{2\pi} e^{il\varphi_r-iQ_{\perp}\rho\cos(\varphi_r-\varphi)}\frac{d\varphi_r}{2\pi}=e^{il\varphi}(-i)^l J_l(Q_{\perp}\rho).
\end{equation}
Following this expression and considering 
asymptotic behavior of the Bessel function, $|J_l(x)|=(x/2)^{|l|}/|l|!$ for $x\rightarrow 0$, we can find that the amplitude
\begin{equation}
\label{eq:Ftom}
    F(\bm{Q},\bm{b}=0)\propto Q_{\perp}^{|l|}\propto(\sin{\theta})^{|l|}.
\end{equation}
When we consider a 
central collision of an 
electron wave packet ($l\neq 0$) with a single atom 
and $\theta=0$, the probability of forward scattering is vanishing. 
However, when $b$ increases, the behavior of the function $U(r)\Psi_{\perp}(\bm{r}+\bm{b})$ changes, and the minimum at $\theta=0$ vanishes. These issues will be 
reckoned with in Sec.~\ref{sec:4}.
\subsection{Scattering on a macroscopic target}
\label{Sec:3D}
Let us now consider 
scattering on a macroscopic (infinitely wide) target with 
randomly distributed potential centers, and with the size of the target 
assumed to be much larger than that 
of the twisted electron wave packet $R\gg \frac{1}{\sigma_{\varkappa}}$. 
To obtain the average 
cross section, we integrate the number of events \eqref{eq:Nu} over all the impact parameters $\bm{b}$ 
and divide this expression by the number of incident electrons $N_e$:
\begin{equation}
\label{eq:SH1}
 \frac{d \Bar{\sigma}}{d\Omega} = \frac{1}{N_e}\int\frac{d \nu}{d\Omega} d^2 \bm{b}.
\end{equation}
When the axis of the wave packet is shifted in the transverse plane by a distance $\bm{b}$ from the potential center, then the corresponding wave function can be written as
\begin{equation}
    \Phi_{\perp}(\bm{k}_{\perp})=a(\bm{k}_{\perp})e^{i\bm{k}_{\perp}\bm{b}},
\end{equation}
where $a(\bm{k}_{\perp})$ is the wave function of a nonshifted packet. Therefore, the expression \eqref{eq:SH1} is proportional to the integral
\begin{equation}
    I=\int F(\bm{Q})F^*(\bm{Q})d^2b
\end{equation}
where
\begin{equation}
    F(\bm{Q})=\int f(\bm{Q}-\bm{k}_{\perp})a(\bm{k}_{\perp})e^{i\bm{k}_{\perp}\bm{b}}\frac{d^2k_{\perp}}{(2\pi)^2}.
\end{equation}
After the integration over $\bm{b}$ and $\bm{k}_{\perp}$, we obtain

\begin{equation}
\label{eq:SH}
    \frac{d \Bar{\sigma}}{d\Omega} = \frac{1}{\cos{\theta_k}}\int |f(\bm{Q}-\bm{k}_{\perp})|^2| \Phi_{\perp}(\bm{k}_{\perp})|^2  d^2 k_{\perp}. 
\end{equation}

In sections Sec.~\ref{Sec:3B} and \ref{Sec:3C} we will 
consider specific different 
central potentials $U(r)$. 
\section{SCATTERING BY HYDROGEN ATOM AND SUPERPOSITION OF THREE YUKAWA POTENTIALS}
\label{Sec:yukawa}
\subsection{Hydrogen atom in the ground state}
\label{Sec:3B}

\begin{table*}
    \begin{ruledtabular}
    \centering
    \begin{tabular}{ p{1.5 cm} | p{1.5 cm} | p{1.5 cm} | p{1.5 cm} |p{1.5 cm} | p{1.5 cm} | p{1.5 cm} | p{1.5 cm}}
        Element & Z & $A_1$ & $A_2$ & $A_3$ & $\mu_1$ & $\mu_2$ & $\mu_3$ \\
        \hline
        Fe & 26 & 0.0512 & 0.6995 & 0.2493 & 31.825 & 3.7716 & 1.1606 \\
         \hline
        Ag & 47 & 0.2562 & 0.6505 & 0.0933 & 15.588 & 2.7412 & 1.1408 \\
         \hline
        Au & 79 & 0.2289 & 0.6114 & 0.1597 & 22.864 & 3.6914 & 1.4886 \\
     
    \end{tabular}
    \end{ruledtabular}
    \caption{Parameters $A_j$ and $\mu_j$ of the Yukawa potential as given in Ref.~\cite{Salvat86,Salvat91}.}
    \label{tab:1}
\end{table*}
Let us consider scattering of a twisted electron wave packet by 
the potential field of a hydrogen atom $U_H(r)$ with an effective action radius $a$
\begin{equation}
U_H(r)=-\frac{e^2}{r}\left(1+\frac{r}{a}\right)e^{-2r/a}
\end{equation}
By inserting this expression into Eq.~\eqref{eq:F} through Eq.~\eqref{eq:f}, we obtain the following plane-wave scattering amplitude 
\begin{equation}
    f_H(\mathbf{q})=\frac{a}{2}\left(\frac{1}{1+(qa/2)^2}+\frac{1}{[1+(qa/2)^2]^2}\right),
\end{equation}
and the scattering amplitude for the twisted electron wave packet will be
\begin{equation}
\label{eq:FH}
  F_H(\bm{Q},\bm{b}) =\frac{a}{2} \int \left(\frac{1}{\omega}+\frac{1}{\omega^2}\right) \Phi_{\perp}(\bm{k}_\perp)e^{i\bm{k}_\perp \bm{b}}\frac{d^2\bm{k_{\perp}}}{2\pi},
\end{equation}
where
\begin{equation}
\label{eq:alphaH}
\begin{aligned}
\omega &=\alpha-\beta \cos(\varphi_k-\varphi), \\
\alpha &= 1+\frac{1}{4} a^2 p_f^2 [1 + \cos^2\theta_k - 2\cos\theta \cos\theta_k]+\frac{1}{4} a^2 k_\perp^2, \\
\beta &= \frac{1}{2} a^2 k_{\perp} p_f \sin\theta.
\end{aligned}
\end{equation}
Eq.~\eqref{eq:FH} can be simplified by introducing the function
   \begin{equation}
   \label{eq:Il}
    I_l(\alpha,\beta, \bm{b})=\int\limits_0^{2\pi}\frac{d\Psi}{2\pi}\frac{e^{il\Psi+ik_{\perp} b \cos{(\Psi+\varphi-\varphi_b)}}}{\alpha-\beta \cos{\Psi}}
\end{equation}
as was proposed in Ref.~\cite{Serbo2015}. 
We can also use the identity~\cite{Karlovets2017}
 \begin{equation}
     \frac{1}{\omega}+\frac{1}{\omega^2}=\left(1-\frac{\partial}{\partial\alpha}\right)\frac{1}{\omega}.
 \end{equation}
 Then Eq.~\eqref{eq:FH} can be rewritten as follows
 \begin{equation}
 \label{eq:FI}
    F_H(\bm{Q},\bm{b}) =\frac{a}{2} \int \limits_0^{\infty} dk_{\perp} \Phi_{tr}(\bm{k}_\perp) \left(1-\frac{\partial}{\partial\alpha}\right) I_l(\alpha,\beta, \bm{b}) .  
 \end{equation}
When we consider 
scattering of an 
electron wave packet on a single hydrogen atom located at the central axis of this packet ($\bm{b}=0$), the function \eqref{eq:Il} becomes 
  \begin{equation}
    I_l(\alpha,\beta, 0)=\left(\frac{\beta}{\alpha+\sqrt{\alpha^2-\beta^2}}\right)^{|l|}\frac{1}{\sqrt{\alpha^2-\beta^2}}.
\end{equation}

Further details are presented in Ref.~\cite{Serbo2015}. From this expression, we can obtain 
the behavior of the angular distribution of the scattered electrons
\begin{equation}
\label{eq:nup}
    \frac{d\nu}{d\Omega}\propto I_l(\alpha,\beta,0)^2\propto\beta^{2|l|}\propto(\sin\theta)^{|l|}.
\end{equation}
When the angle $\theta\rightarrow 0$ 
and $l\neq 0$, the probability of forward scattering disappears. The same result was predicted above when we considered 
scattering by a single atom; see Eq.\eqref{eq:Ftom}.

If an 
electron wave packet collides with an infinitely extending hydrogen target, then, obviously, 
the dependence on the phase 
in Eq.~\eqref{eq:SH} disappears. For a 
BG wave packet dependence on $l$ is exclusively in the phase, and hence its average
cross section is non-
sensitive to 
the OAM. However, for 
LG states \eqref{eq:sLGI}, \eqref{eq:eLGI} there is an additional 
OAM dependence 
in the Laguerre polynomials.
Azimuth part of 
Eq.~\eqref{eq:SH} with the scattering amplitude determined by the potential of the hydrogen atom has the following form: 
\begin{equation}
\begin{aligned}
I_H(\bm{k}_{\perp})&=\frac{a^2}{4}\int\limits_0^{2\pi}\frac{d \varphi_k}{2\pi} \left(\frac{1}{\omega} +\frac{1}{\omega^2}\right)^2 \\
&=\left(-\frac{\partial}{\partial \alpha}+\frac{\partial^2}{\partial \alpha^2}-\frac{1}{6}\frac{\partial^3}{\partial \alpha^3}\right)\frac{1}{\sqrt{\alpha^2-\beta^2}},
\end{aligned}
\end{equation}
where $\omega$, $\alpha$, and $\beta$ are defined in Eq.~\eqref{eq:alphaH}. Then, the average 
cross section is written as
\begin{equation}
      \frac{d \Bar{\sigma}}{d\Omega}=\frac{1}{\cos{\theta_k}}\int \limits_0^{\infty} dk_{\perp} I_{H}(\bm{k}_{\perp})|\Phi_{\perp}(\bm{k}_{\perp})|^2.
\end{equation}

\subsection{Iron, silver, and gold targets}
\label{Sec:3C}
Here we consider iron (Fe), silver (Ag), and gold (Au) targets 
commonly used in experiments. Let us present the potential field as the superposition of three Yukawa terms
\begin{equation}
U_Y(r)=-\frac{Ze^2}{r}\sum_{j=1}^{3}A_j e^{-\mu_jr},
\end{equation}
where  $A_j$ and $ \mu_j$ are determined by the Dirac-Hartree-Fock-Slater 
method~\cite{LIBERMAN1971107,Liberman65}.
Reference numerical values~\cite{Salvat86,Salvat91} are provided in Table~\ref{tab:1}, and $A_3$ is calculated from the condition
\begin{equation}
    A_1+A_2+A_3=1
\end{equation}

Plane-wave scattering amplitude 
determined by Eq.~\eqref{eq:f} 
is written as
 \begin{equation}
f_Y(\mathbf{q})=2 m_e Ze^2\sum_{j=1}^{3}\frac{A_j}{\mathbf{q}^2+\mu_j^2}.
\end{equation}
And the scattering amplitude for the twisted electron has 
the following form
\begin{equation}
\begin{aligned}
  F_Y(\bm{Q},\bm{b}) &=2 m_e Z e^2 \int \left(\sum_{j=1}^{3}\frac{A_j}{\alpha_j-\beta_Y \cos(\varphi_k-\varphi)}\right) \\
 &\times\Phi_{tr}(\bm{k}_\perp)e^{i\bm{k}_\perp \bm{b}}\frac{d^2\bm{k_{\perp}}}{2\pi},
\end{aligned}
\end{equation}
where 
\begin{equation}
\label{eq:alphaY}
\begin{aligned}
&\alpha_j = p_f^2 [1 + \cos^2\theta_k - 2\cos\theta \cos\theta_k]+ k_{\perp}^2+\mu_j^2, \\
&\beta_Y = 2k_{\perp} p_f \sin\theta.
\end{aligned}
\end{equation}
Here, similar to the case of the hydrogen atom, we can rewrite Eq.~\eqref{eq:F} in the form
 \begin{equation}
 \label{eq:FY}
    F_Y(\bm{Q},\bm{b}) =2 m_e Z e^2 \int\limits_0^{\infty} dk_{\perp} \Phi_{\perp}(\bm{k}_\perp)  I_l(\alpha_j,\beta_Y, \bm{b}),
 \end{equation}
 where
    \begin{equation}
   I_l(\alpha_j,\beta_Y, \bm{b})=\int\limits_0^{2\pi}\sum_{j=1}^{3} A_j\frac{e^{il\Psi+ik_{\perp} b \cos{(\Psi+\varphi-\varphi_b)}}}{\alpha_j-\beta_Y \cos{\Psi}}\frac{d\Psi}{2\pi}.
\end{equation} 

If we consider a 
central collision of an 
electron wave packet with a single atom, 
Eq.~\eqref{eq:Il} is rewritten as follows
  \begin{equation}
    I_l(\alpha_j,\beta_Y, 0)=\sum_{j=1}^{3} A_j\left(\frac{\beta_Y}{\alpha_j+\sqrt{\alpha_j^2-\beta_Y^2}}\right)^{|l|}\frac{1}{\sqrt{\alpha_j^2-\beta_Y^2}}.
\end{equation}
Then, if the angle $\theta\rightarrow 0$ and $l\neq0$, we can get 
a result similar to that 
for the hydrogen atom; see Eq.~\eqref{eq:nup}. The angular distribution vanishes for the forward scattering.

When we consider 
scattering of an 
electron wave packet by a macroscopic target, then the azimuth part of 
Eq.~\eqref{eq:SH} with the scattering amplitude determined by the three Yukawa potentials is given as
\begin{equation}
\begin{aligned}
       I_Y&(\bm{k}_{\perp})=4 m_e^2 Z^2e^4 \int\limits_0^{2\pi}\frac{d \varphi_k}{2\pi}\sum_{j=1}^{3} \left(\frac{A_j}{\alpha_j-\beta_Y \cos{(\varphi_k-\varphi)}}\right)^2 \\
        &=\sum_{j=1}^{3}A^2_j\frac{\alpha_j}{\sqrt{(\alpha_j^2-\beta_Y^2)^3}} \\
        &+\sum_{\substack{
  i,j=1 \\
   i\neq j}} ^{3}\frac{A_iA_j \left(\frac{\alpha_i}{\sqrt{\alpha_i^2-\beta_Y^2}}+\frac{\alpha_j} {\sqrt{\alpha_j^2-\beta_Y^2}}\right)}{2\alpha_i \alpha_j-2\beta_Y^2+2\sqrt{\alpha_i^2-\beta_Y^2}\sqrt{\alpha_j^2-\beta_Y^2}} , 
\end{aligned}
\end{equation}
where $\alpha_{i,j}$ and $\beta_Y$ are defined in Eq.~\eqref{eq:alphaY}. Then, the average 
cross section is written as
\begin{equation}
      \frac{d \Bar{\sigma}}{d\Omega}=\frac{1}{\cos{\theta_k}}\int \limits_0^{\infty} dk_{\perp} I_{Y}(\bm{k}_{\perp})|\Phi_{\perp}(\bm{k}_{\perp})|^2.
\end{equation}
\section{RESULTS AND DISCUSSIONS}
\label{sec:4}
Here, we discuss the results of the scattering theory 
developed in Secs.~\ref{Sec:3},~\ref{Sec:yukawa}.
We consider two scenarios: scattering by a single atom 
and scattering by a macroscopic atomic target. 

\subsection{Scattering on a single atom}
 \begin{figure}[t]
\center{\includegraphics[width=1\linewidth]{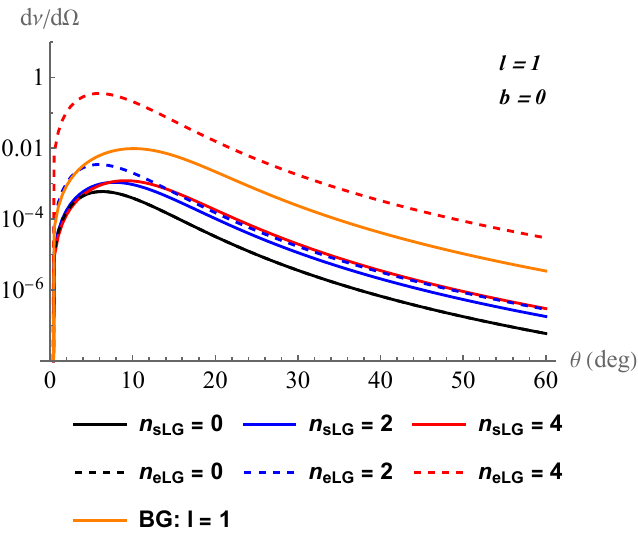}}

  \caption{Scattering of sLG (solid lines), eLG (dashed lines), and BG (orange line) wave packets on a single hydrogen atom: The number of events $d\nu(\theta)/d\Omega$ for 
  different values of $n$, for the atom placed at a distance $b = 0$; $l=1$.}
 \label{fig:SHn}
\end{figure}

The results discussed in this section are comprised in Figs.~\ref{fig:SHn} -- ~\ref{fig:SM} 
where the momentum of the initial electron $p_i=10/ a$ ($\epsilon=1.4$ keV), $a$ being the Bohr radius, the width of the incident wave packets $\sigma_{\varkappa}=\varkappa/5$, opening angle $\theta_k=10^\circ$, and the number of incident electrons $N_e=1$.

\subsubsection{Scattering on a hydrogen atom in the ground 1s state}
We start with scattering of 
electron twisted wave packets on a 
hydrogen atom in the ground $1s$ state.
We present 
the number of events $d\nu(\theta)/d\Omega$~\eqref{eq:Nu} as a function of polar angle $\theta$ for different values of the OAM $l$ and radial number $n$. We also consider two scenarios: when the axis of the incident packet and the center of the target coincide $\bm{b}=0$, and when they are at a distance $\bm{b}$ from each other. 

\begin{figure*}

\begin{subfigure}{0.49\textwidth}
     \includegraphics[width=\textwidth]{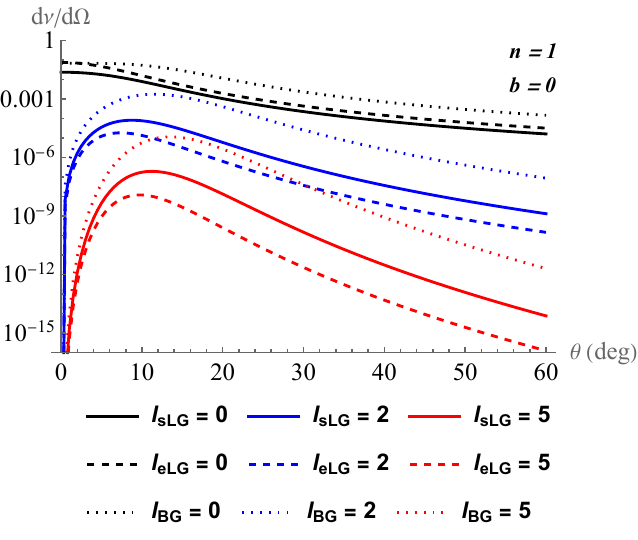}
     \caption{}
 \end{subfigure}
  \hfill
 \begin{subfigure}{0.49\textwidth}
     \includegraphics[width=\textwidth]{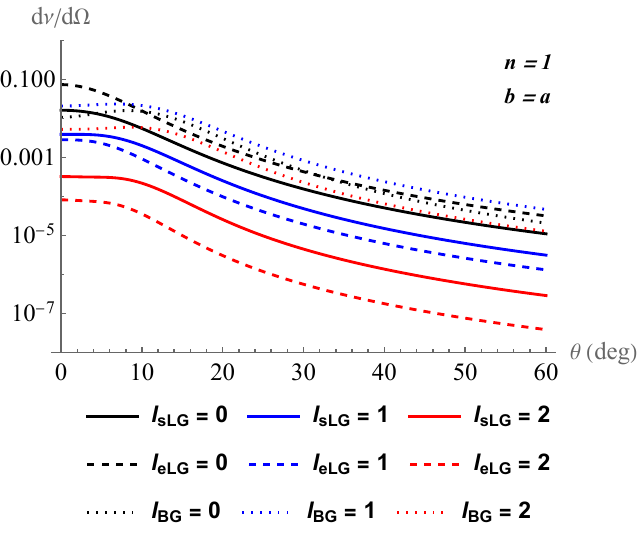}
     \caption{}
      \end{subfigure}
  \caption{Scattering of sLG (solid lines), eLG (dashed lines), and  BG (dotted lines) wave packets on a single hydrogen atom: The number of events $d\nu(\theta)/d\Omega$ for 
  different values of the OAM: $l=0$ (black),$l=2$ (Blue), $l=5$ (Red), for the atom placed at a distance $b = 0$ (a), (c) and $b = a$ (b), (d). The radial quantum number $n=0$ (top panel) and $n = 1$(bottom panel).}
 \label{fig:SHl}
\end{figure*} 

 \begin{figure*}
\begin{subfigure}{0.49\textwidth}
     \includegraphics[width=\textwidth]{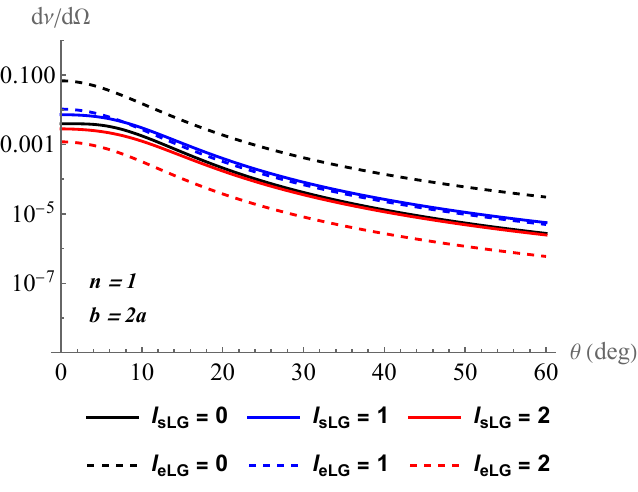}
     \caption{}
 \end{subfigure}
 \hfill
 \begin{subfigure}{0.49\textwidth}
     \includegraphics[width=\textwidth]{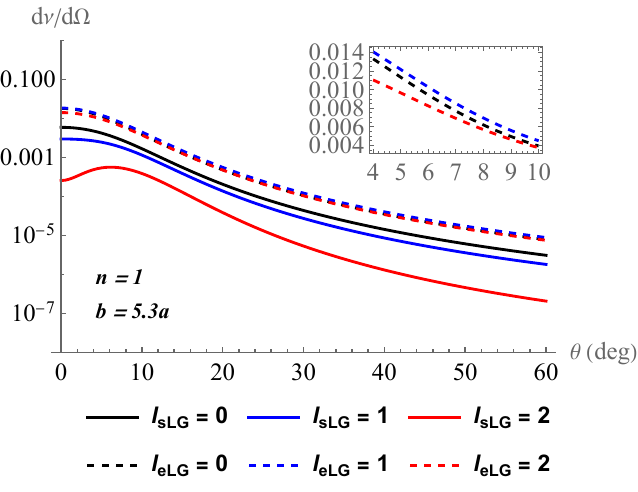}
     \caption{}
      \end{subfigure}
  \caption{Scattering of sLG (solid lines) and eLG (dashed lines) wave packets on a single hydrogen atom: The number of events $d\nu(\theta)/d\Omega$ for 
  different values of $l$, for the atom placed at a distance $b = 2a$ (left panel) and $b = 5.3a$ (right panel), radial quantum number $n=1$.}
 \label{fig:SHb5}
\end{figure*}

Let us 
begin the discussion by considering the number of events for 
twisted wave packets for $l=1$, and for 
different values of $n$: $n=0$ (black), $n=2$ (blue), and $n=4$ (red); see Fig.~\ref{fig:SHn}. When $n=0$, the number of events for 
sLG and eLG wave packets is the same, as anticipated in Sec.~\ref{Sec:2B}.
When $n$ increases, 
the peak in the 
number of events for 
the sLG wave packet shifts towards 
larger values of $\theta$, and the height of the peak also increases, 
approaching the 
number of events for 
the BG wave packet (orange) but always staying beneath it. However, the peak in the number of events for 
the eLG wave packet 
shifts slightly; the number of events for 
$n=4$ is two orders of magnitude larger than that for sLG, and one order of magnitude larger than that for BG packets.
In  Sec.~\ref{Sec:2B}\, when considering Fig.~\ref{fig:Psen}, the spatial distributions of sLG and BG wave packets become similar as the radial number $n$ increases. However, the maximum of the spatial distribution for the BG packet is always higher than for the sLG. Even if we increase $n$ by an order of magnitude, the BG packet has a greater value at the point where the sLG packet has its maximum. Accordingly, the values of the number of events for the sLG packets are always lesser than for the BG packets.
The magnitude of the first peak of the spatial distribution of 
eLG packets significantly increases 
with increasing $n$, which is manifested as 
an increase in the number of events. 
Moreover, with increase of OAM 
the rms radius 
for the sLG packets grows faster than for the eLG ones, as we mentioned in~\ref{Sec:2B}.
Therefore, the maximum number of events for the sLG packets shifts towards larger $\theta$  faster than for the eLG packets. 

In Fig.~\ref{fig:SHl}, we show 
the number of events for 
sLG (solid line), eLG (dashed line), and BG (dotted line) wave packets for three values of OAM: $l=0$ (black), $l=2$ (blue), and $l=5$ (red); the radial number is fixed $n = 1$. We  consider both 
central $\bm{b} = 0$ (left panel) 
and non-central collisions $\bm{b} = a$ (right panel) of the wave packets with the atomic target.
As can be seen from Fig.~(\ref{fig:SHl},a), 
for a 
central collision, 
the number of events for 
eLG and sLG wave packets differ, and their divergence becomes even more pronounced 
with increasing $l$.
While for $l=0$, the number of events for 
the eLG packet is greater than that of 
the sLG packet, for other OAM values, we have the opposite result. Figure~\ref{fig:PseL} shows that for $l=0$, the peak value of the spatial distribution  of the eLG packet is 
greater than that of the sLG packet.
However, for larger $l$, the first ring of the
sLG packet is 
closer to the center of the target than that of the
eLG packet. Therefore, a larger part of the
sLG packet is scattered on the target, and hence the values of the number of events are larger than those for the
eLG packet.

Furthermore, in Fig.~\ref{fig:SHl} 
the  maximum of the number of events for 
sLG and eLG wave packets is lower 
than that for the BG wave packet -- the peak of the spatial distribution of a BG wave packet is located closer to the center of the target, 
as seen in Fig.~\ref{fig:PseL}.
Therefore, a larger part of the BG packet hits the target, and we observe larger values for the corresponding number of events. 

Additionally, in Fig.~(\ref{fig:SHl},a), 
we observe a dip in the forward scattering of the electrons for $l>0$ and $b=0$ for all the twisted wave packets, following Eqs.~\eqref{eq:Ftom},~\eqref{eq:nup}; however, for
finite values of $b$ (Fig.~(\ref{fig:SHl},b)) 
forward scattering becomes 
allowed, as discussed in Secs.~\ref{Sec:3A},~\ref{Sec:3B}.

For the non-central collision shown in Fig.~(\ref{fig:SHl}, b), the relative position of the number of events curves for sLG and eLG packets is the same as in Fig.~(\ref{fig:SHl},a) for the central collision, and the previous explanation remains. However, here we observe that the values for eLG with $l=0$ near small $\theta$ are larger than even those for the BG packets. We can assume that in the position of displacement by the impact parameter $b=a$ the spatial distribution of the BG packet with $l=0$ is close to its first minimum (recall Fig.~(\ref{fig:PseL}, a)), whereas the distribution values for the eLG packet are still far from the minimum and several orders of magnitude higher.

Fig.~(\ref{fig:SHl}, b) also shows that for 
sLG and eLG wave packets, the highest values of the number of events 
are observed for $l=0$, and for the BG one,  
they are observed for $l=1$, 
because the first ring of the BG packet is located at the target axis.
As can be seen in Fig.~(\ref{fig:PseL}, b), the first ring of 
sLG and eLG packets is located further 
from the target center than that of 
the BG ring. 
At the point where the BG wave packet has a maximum  for $l=1$, the spatial distributions of 
sLG and eLG packets have larger 
magnitudes for $l = 0$ and lesser for $l=1$. Accordingly,
the highest-lying curves of the number of events for sLG and eLG wave packets correspond to $l=0$.

Let us investigate when the highest values of the number of events are achieved for $l\neq 0$ in case of sLG and eLG wave packets.
If we increase the impact parameter to $b=2a$, then for 
sLG wave packets the highest values of the number of events correspond to $l=1$, as 
can be seen in Fig.~(\ref{fig:SHb5}, a).
If the impact parameter is equal to $b=5.3a$, then the same effect is observed for the
eLG wave packet collision, as 
shown in Fig.~(\ref{fig:SHb5}, b). Here, the curves of the number of events for 
eLG packets with different $l$ are 
almost identical, but the highest-lying one corresponds to $l=1$.
We can conclude that the bigger the radius of the first ring of a twisted packet is, the larger 
impact parameter $b$ is required for the values of the number of events for an electron wave packet with $l=1$ to become dominant. We  note that with a further increase in $b$, the highest values of the number of events correspond to 
wave packets with larger OAM values ($l=2, 3, \dots$). 

Using the obtained results, 
we can perform quantum tomography of 
wave packets by shifting the target by a certain impact parameter $b$, thus using the atom as a probing tool. However, from the experimental point of view, setting the impact parameter with such accuracy is very challenging,
and the difference in the scattering patterns will be difficult to see.

\subsubsection{Scattering on Fe, Ag, and Au atoms}
\label{Sec:32}
\begin{figure*}
 \begin{subfigure}{0.32\textwidth}
     \includegraphics[width=\textwidth]{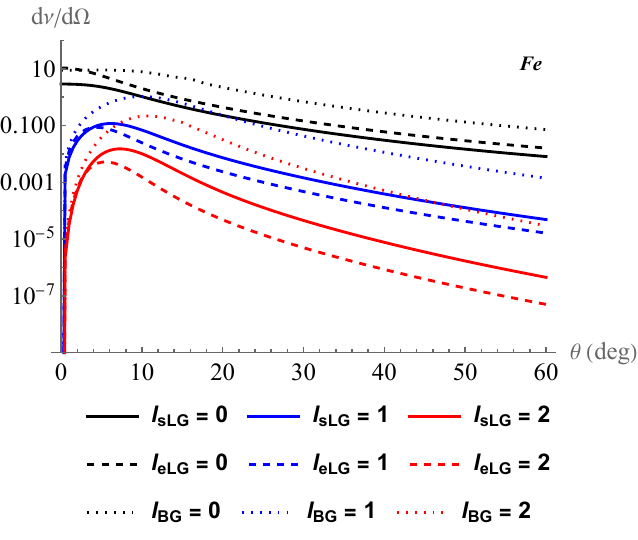}
     \caption{Fe}
 \end{subfigure}
 \hfill
 \begin{subfigure}{0.32\textwidth}
     \includegraphics[width=\textwidth]{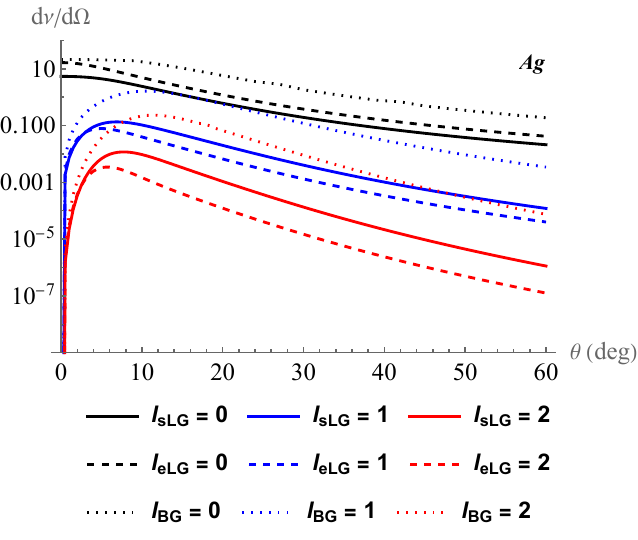}
     \caption{Ag}
 \end{subfigure}
\hfill
 \begin{subfigure}{0.32\textwidth}
     \includegraphics[width=\textwidth]{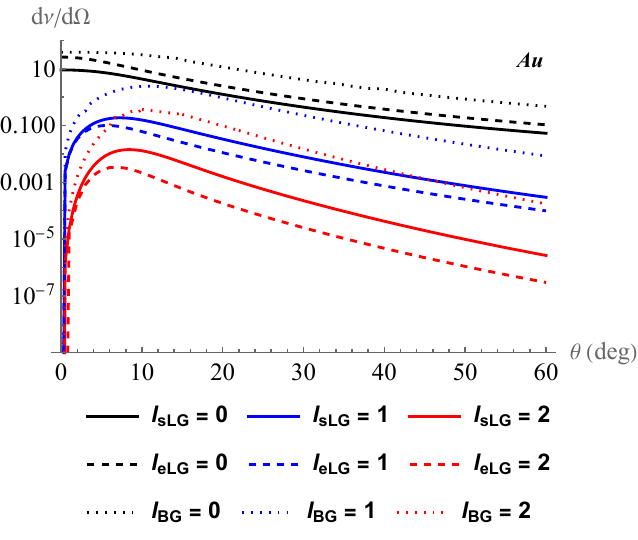}
     \caption{Au}
 \end{subfigure}
  \caption{Scattering of 
  sLG (solid lines), eLG (dashed lines), and  BG (dotted lines) packets on Fe (a), Ag (b), and Au (c) atoms: The number of events $d\nu(\theta)/d\Omega$ for 
  different values of the OAM: $l=0$ (black), $l=1$ (Blue), $l=2$ (Red), for the atom placed at a distance $b = 0$. The width of the incident packets is $\sigma=\varkappa/5$, the opening angle $\theta_k=10^\circ$.}
 \label{fig:SM}
\end{figure*}

\begin{figure*}
 \begin{subfigure}{0.32\textwidth}
     \includegraphics[width=\textwidth]{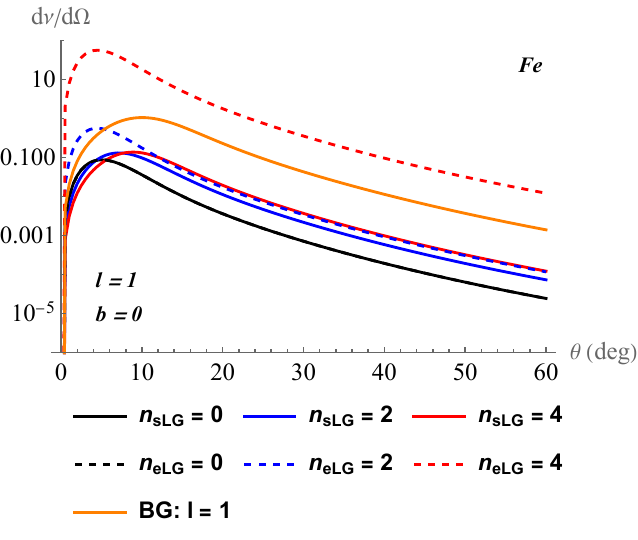}
     \caption{}
 \end{subfigure}
\hfill
 \begin{subfigure}{0.32\textwidth}
     \includegraphics[width=\textwidth]{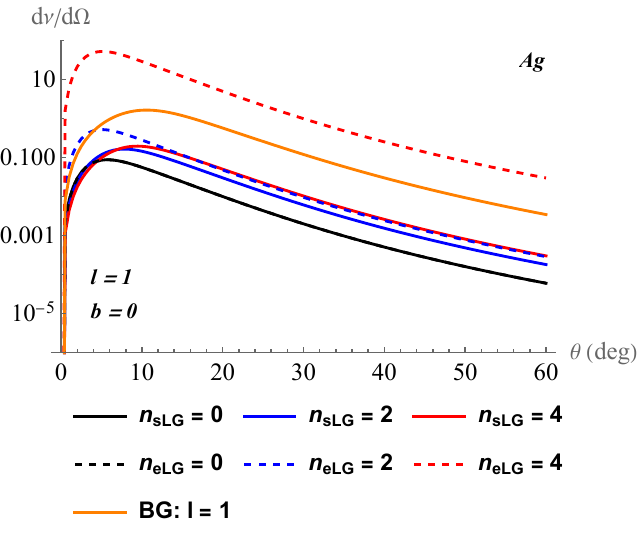}
     \caption{}
 \end{subfigure}
\hfill
 \begin{subfigure}{0.32\textwidth}
     \includegraphics[width=\textwidth]{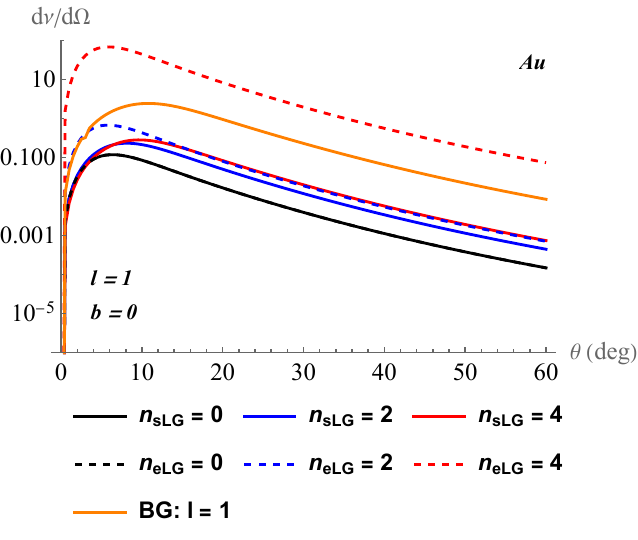}
     \caption{}
 \end{subfigure}
 
  \caption{Scattering of 
  sLG (solid lines), eLG (dashed lines), and  B
  G (orange line) packets on Fe (a), Ag (b), and Au (c) atoms: The number of events $d\nu(\theta)/d\Omega$ for 
  different values of $n$ for the atom placed at a distance $b = 0$. The width of the incident packets is $\sigma=\varkappa/5$, the opening angle $\theta_k=10^\circ$, and the OAM $l=1$.}
 \label{fig:SYn}
\end{figure*}


Further, we consider 
central collisions of 
twisted wave packets with 
metal targets. Figure \ref{fig:SM} presents
the number of events for electron sLG (solid line), eLG (dashed line), and BG (dotted line) wave packets with $n=1$ and different values of $l$: $l=0$ (black), $l=1$ (blue), and $l=2$ (red) scattered on iron (Fig.~(\ref{fig:SM},a)), silver (Fig.~(\ref{fig:SM},b)), and gold (Fig.~(\ref{fig:SM},c)) targets.

We can see that the scattering patterns for 
twisted packets interacting with a 
hydrogen atom and with 
metal targets are similar.
However, here the values of the number of events is two orders of magnitude higher 
than those for a hydrogen atom. The reason is, 
the number of events depends on the nuclear charge $Z^2$, see Eqs.~(\ref{eq:Nu},~\ref{eq:FY}). Notably, with increasing 
$Z$, the slope 
of the number of events 
on 
the right side of 
the peak becomes less steep.
This 
can be seen from Eq.~\eqref{eq:FY}, 
where $I_l(\alpha_j,\beta, \bm{b}=0)$ decreases when an element with larger $Z$ is chosen, but if we take into account the prefactor $2m_eZe^2$ in Eq.~\eqref{eq:FY}, the overall amplitude $F_Y(\bm{Q},\bm{b})$ grows with $Z$.  

In Fig.~\ref{fig:SYn}, we show 
$d\nu(\theta)/d\Omega$ for LG wave packets with different values of $n$: $n = 0$ (black line), $n=2$ (blue line), and $n=4$ (red line) scattered on iron (Fig.~(\ref{fig:SYn},a)), silver (Fig.~(\ref{fig:SYn},b)), and gold (Fig.~(\ref{fig:SYn},c)) targets. The OAM value is fixed to $l = 1$ and scattering of the BG wave packet 
is indicated with an orange line. The distinction in the overall scale of the number of events between hydrogen and heavy elements that we observed in Fig.~(\ref{fig:SM}) is preserved when the radial number $n$ is varied.
Here, as for the scattering on the hydrogen atom,  the pattern of the sLG wave packet scattering becomes similar to that 
of the BG wave packet with an increase of $n$. However, this is not the case 
for the eLG wave packet, for which 
with an increase of $n$, the value of the maximum of the number of events becomes larger than 
for the BG packet.

\subsection{Macroscopic target}
\begin{figure*}
 \begin{subfigure}{0.49\textwidth}
     \includegraphics[width=\textwidth]{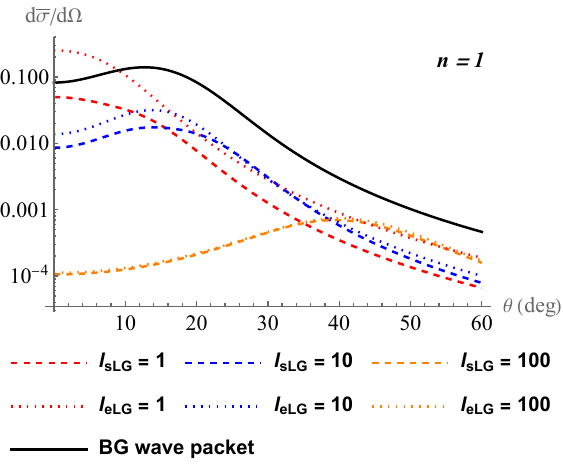}
   \caption{}
\end{subfigure}
 \hfill
 \begin{subfigure}{0.49\textwidth}
     \includegraphics[width=\textwidth]{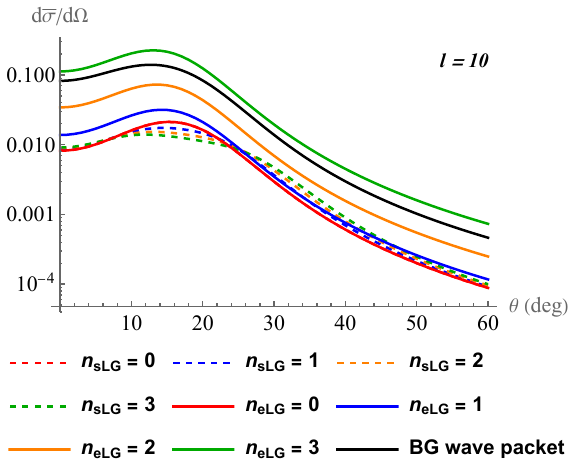}
     \caption{}
 \end{subfigure}
 
  \caption{Scattering on macroscopic target: The average 
  cross section $d\Bar{\sigma}/d\Omega$ for 
  different values of OAM (a) and $n$ (b). Results are presented for the width of the incident packet $\sigma_{\varkappa}=\varkappa_0/3$ and opening angle $\theta_k=15^\circ$. Left panel corresponds to $n = 1$ and right panel to $l = 10$.}
 \label{fig:Macro}
\end{figure*}

Now we consider 
scattering of 
wave packets on a macroscopic (infinite-size) atomic target. 
Similar to 
the previous section, we focus on LG wave packets with $n > 0$, because for $n = 0$, the average 
cross sections for 
the sLG and eLG wave packets coincide. 
Figures (\ref{fig:Macro}, a),~(\ref{fig:Macro}, b) show 
the average 
cross sections $d\Bar{\sigma}/d\Omega$ for different values of OAM $l$ with fixed $n = 1$, and for different values of quantum number $n$ with fixed $l = 10$; 
the width of the incident packets is $\sigma_{\varkappa}=\varkappa_0/3$ and the opening angle $\theta_k=15^\circ$.

We observe that $d\Bar{\sigma}/d\Omega$ for 
sLG and eLG packets is sensitive to the OAM, in a sense that one can easily distinguish single-digit values of OAM from those counting dozens or hundreds of $l$. In Fig.~(\ref{fig:Macro}, a), we plot the average cross sections for $l=1$ (red line), $l=10$ (blue line), and $l =100$ (orange line). Note that for the BG packet (black line), the dependence on $l$ is 
absent.
It follows from the presence of OAM in the Laguerre polynomial in Eq.~\eqref{eq:SH}, as discussed in Sec.~\ref{Sec:3D}. Besides, with an 
increase of $l$, the maximum of the average 
cross sections shifts towards
larger $\theta$ values and becomes lower. 
This happens because the rms radii squared of the LG packets are proportional to $n$ and $l$ \cite{NJP2021}, and grow with increasing OAM, whereas the values of the first peak decrease.


\begin{figure*}
 \begin{subfigure}{0.32\textwidth}
     \includegraphics[width=\textwidth]{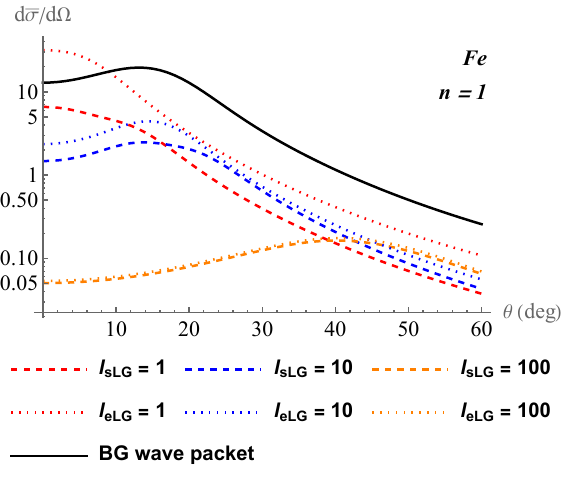}
     \caption{}
 \end{subfigure}
\hfill
 \begin{subfigure}{0.32\textwidth}
     \includegraphics[width=\textwidth]{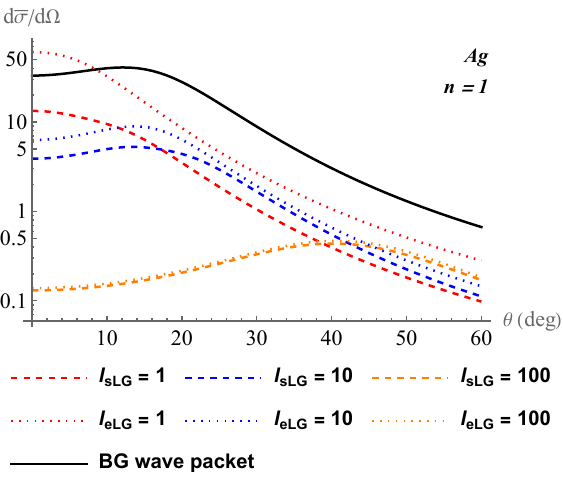}
     \caption{}
 \end{subfigure}
\hfill
 \begin{subfigure}{0.32\textwidth}
     \includegraphics[width=\textwidth]{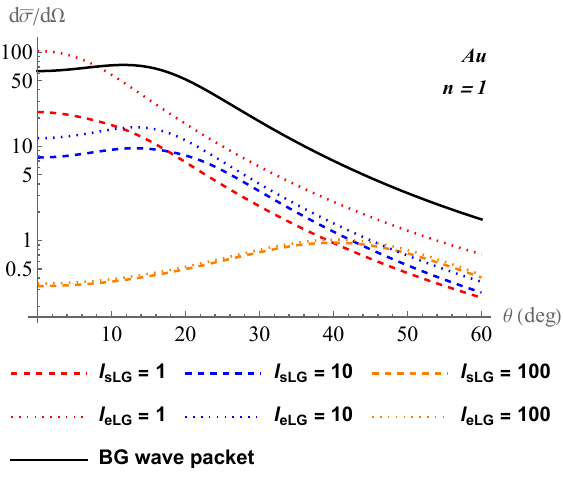}
     \caption{}
 \end{subfigure}
 
  \caption{Scattering of 
  sLG (solid lines), eLG (dashed lines), and  BG (orange line) packets on Fe (a), Ag (b), and Au (c) macroscopic targets: The average 
  cross section $d\Bar{\sigma}/d\Omega$ for 
  different values of $l$. Width of the incident packets $\sigma=\varkappa/3$, opening angle $\theta_k=15^\circ$, and $n=1$.}
 \label{fig:MY}
\end{figure*}

In Fig.~(\ref{fig:Macro}, a), we can see that the values of the average cross section for sLG and eLG wave packets are more distinguishable for smaller OAM values. The reason is, 
the magnitude of the peak of the spatial distribution of 
an eLG packet is much larger than that of an sLG packet; see Fig.~\ref{fig:PseL}. However, for larger OAM values, 
for example, $l = 100$, the cross sections are 
almost identical, because for $n=1$ 
and large $l$ the difference between the wave packets \eqref{eq:sLGI} and \eqref{eq:eLGI} is small. On the other hand, for $n\ll l$, 
the polynomials in the sLG and eLG packets become increasingly similar. Should we consider a greater value of the radial number, the difference in the scattering patterns of 
sLG and eLG packets would be pronounced for a wider range of $l$, but still would disappear for $l \gg n$. 

Additionally, 
when a BG wave packet is scattered on a macroscopic target, its scattering pattern is determined by the width of the packet, see details in Ref.~\cite{Karlovets2017}. 
Let us take the same width of the twisted packets $\sigma_{\varkappa}=\varkappa/3$. Then, we note that it is for $l = 10$ (blue lines in Fig.~(\ref{fig:Macro}, a)) when the peaks for BG and LG packets are situated around the same $\theta$ values and, thus, scattering patterns are analogous. However, if we want to achieve the closest numerical values of the averaged cross sections for the LG and BG packets, taking LG packets with $l=1$ seems a better option.
We conclude that in experimental setups sLG and eLG packets could be distinguished from BG ones by studying the sensitivity of the average cross section to OAM akin to what is depicted in Fig.~(\ref{fig:Macro}, a).

At the same time, Fig.~(\ref{fig:Macro}, b) shows the average cross sections for different values of $n$. 
Changing $n$ does not influence 
the average cross section for the
sLG wave packet. However, with an
increase of $n$, the values of the average cross section for the
eLG wave packets increase by an order of magnitude. That is not surprising because the peak of the spatial distribution of this packet rises with growing $n$ as could be seen in Fig.~\ref{fig:Psen}.
Currently, 
the difference between sLG and eLG packets cannot be revealed in experiment by changing the radial quantum number $n$, because one cannot specify 
$n$ in practice. However, if it becomes possible, 
this difference will be visible when $n$ increases from one to four (with other wave packet parameters set as in Fig.~\ref{fig:Macro}), since in this case, the average cross section for 
the eLG packet increases by a whole order of magnitude, whereas that of 
the sLG packet 
does not change.

In Fig.~\ref{fig:MY}, we present the same patterns as in Fig.~\ref{fig:Macro}, but 
here, the wave packets are scattered on metal (Fe, Ag, and Au) targets. 
We can see that for the scattering on a macroscopic target made of metal the values scale up compared to the target with the Hydrogen atoms (Fig.~(\ref{fig:SM})). Also, the maximum of the cross section flattens out with increasing $Z$. Overall, the behavior is similar to the one discussed in Sec.~\ref{Sec:32}. 
Notably, for macroscopically distributed heavy atoms the scale of the cross section keeps growing for larger values of $Z$. This is in contrast to what is observed in the single atom scenario. Larger numerical values of $d\bar{\sigma}/d\Omega$ for heavier $Z$ permit greater partition of scattering data for wave packets with smaller values of OAM. This allows one to better distinguish different wave packets in the experiment.
\section{CONCLUSION}
\label{Sec:5}

We have developed the theory of elastic scattering of 
well-localized twisted packets by atoms in the framework of Born approximation. We have considered 
the following scenarios: scattering twisted packets by a single atom and by a macroscopic target.  

For the scattering on a single atom we reveal 
that the number of events depends on the form of the incident packet and the location of the target, determined by the impact parameter $b$. Since the first ring of a BG 
packet is smaller than that of a sLG 
packet, the number of events for BG packets centrally colliding with the target achieves larger values than that for the sLG packet.
The scattering processes for the eLG 
and sLG wave packets are the same when $n=0$. However, for any given set of wave packets parameters, with an increase of $n$, the values of the number of events for 
eLG packets would eventually become substantially larger than those for sLG packets. With $l=1$ and other parameters considered in this paper, for the radial number $n=2$ we still observe the similarity in the number of events for the eLG and sLG scenarios, whereas for $n=4$, the values of the number of events for the eLG packet exceed the sLG ones by approximately four orders of magnitude and the BG ones by two orders. Were it possible to have a good control over $n$ in experiment, this would have given an opportunity to distinguish the type of the generated LG packet.
When we consider heavier atoms as targets, the overall scale of the number of events increases by two orders of magnitude relative to the values characteristic to the scattering by the hydrogen atom.

We also consider the scattering pattern for 
different values of the impact parameter $b$. 
 We show that it is possible to find such an impact parameter that the values of the number of events for packets with $l = 1$ surpass the values for the packets of the same type with other $l$. 
 For BG packet, 
this effect appears at $b=a$, for sLG packet, at $b=2a$, and for 
eLG packet, at $b=5.3a$ (we took the radial number $n=1$ for Laguerre - Gaussian packets). With a further increase in the impact parameter, the essence of the effect remains but the highest values are achieved for $l=2$ and so on. 
That is by shifting the target away from the center of the incident packet, it could be possible to determine from the scattering pattern which type of the wave packet was generated. 
Alternatively, if the experimenter is confident in the type of the wave packet, studying the scattering patterns at specific target locations provides information on the quantum numbers of the wave packets, i.e, it is akin to performing quantum tomography of wave packets using an atom as a probing device. 
However, we acknowledge that at the moment it is quite difficult to achieve such experimental accuracy in setting the impact parameter $b$. 



In case of 
scattering of 
electron sLG and eLG wave packets by 
a macroscopic target, 
the values of the average 
cross section vary substantially when the OAM values of the incident wave packet are changed by an order of magnitude. This fact will allow us to experimentally specify the OAM order with great accuracy. 
There are also differences between the values for sLG and eLG packets, although they 
can be seen only for small values of OAM, since when $l \gg n$ these packets are virtually the same. 
The values of the average 
cross section for 
the sLG packets have a low sensitivity 
to variations in $n$, when the opposite is observed for eLG packets
: 
the higher $n$ is, the larger the value of the average 
cross section becomes. 

When we consider 
scattering on an 
metal target, the values of the average cross section increase greatly with element charge $Z$. There is also an increase in the difference between the values of the average cross sections for 
sLG and eLG packets with 
small values of $l$. 
From the experimental point of view, it 
 means that a macroscopic target consisting of atoms with larger $Z$ enables better differentiation between the types of twisted packets from the result of their scattering.
 

\section*{ACKNOWLEDGMENTS}
The work  on the quantum states (by D. Karlovets and D. Grosman) in Sec.II  was supported by the Foundation for the Advancement of Theoretical Physics and Mathematics “BASIS”.  The studies in Sec. III are supported by the Government of the Russian Federation through the ITMO Fellowship and Professorship Program.
The studies in Sec. IV are supported by the Russian Science Foundation (Project No.\,23-62-10026;  https://rscf.ru/en/project/23-62-10026/).
{The studies in Sec.V are supported by the Ministry
of Science and Higher Education of the Russian Federation (agreement No. 075-15-2021-1349).}

\newpage
\bibliographystyle{apsrev}
\bibliography{LGB}
\end{document}